\documentclass[12pt]{article}
\usepackage{amsfonts}
\usepackage{epsfig,amssymb,euscript}
\usepackage{amsmath,amscd}

\numberwithin{equation}{section}
\newcommand{\be}{\begin{eqnarray}}
\newcommand{\ee}{\end{eqnarray}}
\newcommand{\bea}{\begin{eqnarray}}
\newcommand{\eea}{\end{eqnarray}}
\newcommand{\ba}{\begin{array}}
\newcommand{\ea}{\end{array}}

\begin{document}

\begin{titlepage}
\vfill
\begin{flushright}
\end{flushright}
\vfill
\begin{center}
   \baselineskip=16pt
  {\Large\bf HKT Geometry and Fake Five Dimensional Supergravity}
   \vskip 2cm
       Jan B. Gutowski$^1$, 
      and W. A. Sabra$^2$\\
 \vskip .6cm
      \begin{small}
      $^1$\textit{Department of Mathematics, King's College London.\\
      Strand, London WC2R 2LS\\United Kingdom \\
        E-mail: jan.gutowski@kcl.ac.uk}
        \end{small}\\*[.6cm]
  \begin{small}
      $^2$\textit{Centre for Advanced Mathematical Sciences and
        Physics Department, \\
        American University of Beirut, Lebanon \\
        E-mail: ws00@aub.edu.lb}
        \end{small}      
   \end{center}
\vfill
\begin{center}
\textbf{Abstract}
\end{center}
\begin{quote}
Recent results on the relation between hyper-K\"ahler geometry with Torsion and 
solutions admitting Killing spinors in minimal de sitter supergravity  are extended to more general 
supergravity models with vector multiplets. 
\end{quote}
\end{titlepage}%

\section{Introduction}

The strong relation between complex geometry and supersymmetry has been
known for some time by now. It was observed first by Zumino 
\cite{Zumino:1979et} that demanding $N=2$ supersymmetry on a two-dimensional
non-linear sigma model puts the restriction that the target space metric
must be described by a K\"{a}hler manifold. Extending the supersymmetry to 
$N=4$, the target manifold then becomes a hyper-K\"{a}hler manifold 
\cite{AlvarezGaume:1981hm}. The Wess-Zumino-Witten couplings \cite{Witten:1983ar}
in the non-linear sigma model can be interpreted as torsion potentials from
the target space viewpoint \cite{Curtright:1984dz,Braaten:1985is}. Thus, it
is natural to expect that adding such couplings will lead to K\"{a}hler and
hyper-K\"{a}hler torsion (HKT) target space geometries \cite{howepaphkt,
hullhkt, hull:1984, hull:1986}.

In string compactifications, demanding that the four dimensional low energy
action has $N=1$ supersymmetry forces the six dimensional compact manifold
to be a Calabi-Yau 3-fold \cite{Candelas:1985en}. Another connection between
complex geometry and supersymmetry was also revealed in the study of the
moduli space metric of supersymmetric electrically charged five-dimensional
black holes which was found to be described by a HKT\ manifold 
\cite{gibbonshkt, multihkt}.

More recently, K\"{a}hler and hyper-K\"{a}hler geometry also arise in
connection with the study of supersymmetric solutions in supergravity
theories. For instance, the four-dimensional base spaces of time-like
supersymmetric solutions of ungauged and gauged five dimensional
supergravity are given, respectively, by a hyper-K\"{a}hler and K\"{a}hler
manifold \cite{Gauntlett:2002nw, Gauntlett:2003fk}.

The embedding of cosmological Einstein gravity in a supergravity theory is
allowed provided that the cosmological constant is either vanishing or
negative. However, in the case of a positive cosmological constant, the
concept of fake supergravity can be introduced as a solution generating
technique. In this case, a Killing spinor equation is obtained from the analytic
continuation of the equation resulting from the vanishing of the gravitini
supersymmetry variation in the corresponding theory with negative
cosmological constant. Recently, the programme of the classification of all
solutions admitting (pseudo-)Killing spinors in de Sitter supergravity
theories was initiated in \cite{hkt5}. There it was shown that the base
space of time-like solutions of five dimensional de Sitter supergravity is
given by four dimensional HKT\ geometry. Moreover, solutions admitting null
Killing vectors were later analysed in \cite{gtnull5d} where it was found
that those solutions are related to a one-parameter family of Gauduchon-Tod
spaces \cite{gaudtod}. Our present work is the generalisation of the results
of \cite{hkt5} to five dimensional supergravity models with scalar fields
which could be of relevance to cosmological models.

This paper is organized as follows. Section two contains a brief description
of the models under study and the analysis of the fake gravitino and
dilatino Killing spinor equations of the five dimensional de Sitter
supergravity with vector multiplets. The general structure of the
pseudo-supersymmetric solutions, admitting Killing spinors that give rise to
a timelike vector field, is obtained. In section three and four we provide
some examples and in section five we give some final remarks.

\textit{Note added}: At the time of completion of this work, we have become
aware of the work of \cite{maeda}, in which a subclass of the 
pseudo-supersymmetric solutions with a hyper-K\"{a}hler base space was
examined. However, in general the base space is hyper-K\"{a}hler with torsion, and the
classification constructed in this paper describes the most general
pseudo-supersymmetric solution.

\section{Fake $N=2$ supergravity and Killing Spinors}

The model we will be considering in our present work is $N=2$, $D=5$
gauged supergravity coupled to abelian vector multiplets \cite{gunaydin}
whose bosonic action is given by 
\begin{equation}
S={\frac{1}{16\pi G}}\int \left( R+2g^{2}{\mathcal{V}}\right) 
{\mathcal{\ast }}1-Q_{IJ}\left( dX^{I}\wedge \star dX^{J}+F^{I}\wedge \ast
F^{J}\right) -{\frac{C_{IJK}}{6}}F^{I}\wedge F^{J}\wedge A^{K}
\label{action}
\end{equation}
where $I,J,K$ take values $1,\ldots ,n$ and $F^{I}=dA^{I}$ are the two-forms
representing gauge field strengths (one of the gauge fields corresponds to
the graviphoton). The constants $C_{IJK}$ are symmetric in $IJK$, we will
assume that $Q_{IJ}$ is invertible, with inverse $Q^{IJ}$. The $X^{I}$ are
scalar fields subject to the constraint 
\begin{equation}
{\frac{1}{6}}C_{IJK}X^{I}X^{J}X^{K}=X_{I}X^{I}=1\,.  \label{eqn:conda}
\end{equation}
The fields $X^{I}$ can thus be regarded as being functions of $n-1$
unconstrained scalars $\phi ^{r}$. We list some useful relations 
\begin{eqnarray}
Q_{IJ} &=&{\frac{9}{2}}X_{I}X_{J}-{\frac{1}{2}}C_{IJK}X^{K} \nonumber  \\
\text{ \ \ \ }Q_{IJ}X^{J} &=&{\frac{3}{2}}X_{I}\,,\qquad Q_{IJ}dX^{J}=
-{\frac{3}{2}}dX_{I}\ , \nonumber \\
{\mathcal{V}} &=&9V_{I}V_{J}(X^{I}X^{J}-{\frac{1}{2}}Q^{IJ}) \ ,
\end{eqnarray}
here $V_{I}$ are constants.

Fake supergravity theory is obtained by sending $g^{2}$ to $-g^{2}$ in the
above action. We start out analysis of pseudo-supersymmetric de-Sitter
solutions by examining the fake gravitino Killing spinor equation:

\begin{equation}
\left[ \nabla _{M}-{\frac{i}{8}}\Gamma _{M}H_{N_{1}N_{2}}\Gamma
^{N_{1}N_{2}}+{\frac{3i}{4}}H_{M}{}^{N}\Gamma _{N}-
g(\frac{i}{2}X\Gamma _{M}-\frac{3}{2}A{}_{M})\right] \epsilon =0,  \label{fakekilling}
\end{equation}
where we have defined 
\begin{equation}
V_{I}X^{I}=X,\text{ \ \ \ \ \ }V_{I}A^{I}{}_{M}=A{}_{M},\text{ \ \ \ \ \ \ }
X_{I}F^{I}{}_{MN}=H_{MN}.
\end{equation}

We shall analyse the solutions of the Killing spinor equations using spinorial
geometry techniques originally developed to analyse supersymmetric solutions in
ten and eleven dimensional supergravity \cite{papadopd11, papadopiib}, and
which have since been used to analyse a large variety of supersymmetric solutions
in numerous theories.
For de Sitter supergravity in five-dimensions, one takes the space of Dirac
spinors to be the space of complexified forms on $\mathbb{R}^{2}$, which are
spanned over $\mathbb{C}$ by $\{1,e_{1},e_{2},e_{12}\}$ where 
$e_{12}=e_{1}\wedge e_{2}$. The action of complexified $\Gamma $-matrices on
these spinors is given by

\begin{eqnarray}
\Gamma _{\alpha } &=&\sqrt{2}e_{\alpha }\wedge \ , \nonumber \\
\Gamma _{\bar{\alpha}} &=&\sqrt{2}i_{e_{\alpha }}\ ,
\end{eqnarray}
for $\alpha =1,2$, and $\Gamma _{0}$ satisfies 
\begin{equation}
\Gamma _{0}1=-i1,\quad \Gamma _{0}e^{12}=-ie^{12},\quad \Gamma
_{0}e^{j}=ie^{j}\ \ j=1,2\ .
\end{equation}

The spacetime metric has signature $(-,+,+,+,+)$ and is written in the
following basis 
\begin{equation}
ds^{2}=-(\mathbf{e}^{0})^{2}+2\delta _{\alpha \bar{\beta}}\mathbf{e}^{\alpha
}\mathbf{e}^{\bar{\beta}}\ .  \label{metricform}
\end{equation}
The $Spin(4,1)$ gauge transformations can be used to fix the Killing spinor
to take the form $\epsilon =f1$. Moreover, we can set $f=1,$ using the 
$\mathbb{R}$ transformation \cite{hkt5} 

\begin{equation}
\epsilon \rightarrow e^{\lambda }\epsilon ,\qquad V_{I}A^{I}\rightarrow
V_{I}A^{I}-{\frac{2}{3g}}d\lambda ,
\end{equation}
which leaves the Killing spinor equation invariant$.$ With all this
information, we obtain from (\ref{fakekilling}), the following conditions:

\begin{eqnarray}
H_{\alpha }^{\text{ \ }\alpha }+2{g}X-6{g}A{}_{0}
-2\Omega _{0,\alpha }^{\text{ \ \ \ \ \ }\alpha } &=&0,  \notag \\
H_{0\alpha }-\Omega _{0,0\alpha } &=&0,  \notag \\
\left( \Omega _{0,\bar{\alpha}\bar{\beta}}
-{\frac{1}{2}}H_{\bar{\alpha}\bar{\beta}}\right) \epsilon ^{\bar{\alpha}\bar{\beta}} &=&0,  \notag \\
\frac{1}{2}\Omega _{\beta ,\alpha }^{\text{ \ \ \ \ }\alpha }
+{\frac{3}{4}}H_{0\beta }+\frac{3g}{2}A{}_{\beta } &=&0,  \notag \\
\Omega _{\alpha ,0\bar{\beta}}+\frac{1}{2}H_{\mu }^{\text{ \ }\mu }\delta
_{\alpha \bar{\beta}}-{\frac{3}{2}}H_{\alpha \bar{\beta}}+{g}X\delta
_{\alpha \bar{\beta}} &=&0,  \notag \\
\Omega _{\beta ,\bar{\mu}\bar{\nu}}\epsilon ^{\bar{\mu}\bar{\nu}}+H^{0\mu
}\epsilon _{\alpha \mu } &=&0,  \notag \\
\Omega _{\bar{\alpha},0\bar{\beta}}-\frac{1}{2}H_{\bar{\alpha}\bar{\beta}}
&=&0,  \notag \\
\Omega _{\bar{\beta},\mu }{}^\mu+\frac{1}{2}H_{0\bar{\beta}}
+3gA{}_{\bar{\beta}} &=&0,  \notag \\
\Omega _{\bar{\beta},\bar{\mu}\bar{\nu}}\epsilon ^{\bar{\mu}\bar{\nu}} &=&0.
\label{gravitini}
\end{eqnarray}
The above equations then imply:

\begin{eqnarray}
A{}_{0} &=&\frac{X}{3},  \notag \\
A{}_{\alpha } &=&-\frac{1}{3g}\Omega _{0,0\alpha },
\notag \\
H_{0\alpha } &=&\Omega _{0,0\alpha },  \notag \\
H_{\alpha \bar{\beta}} &=&\frac{2}{3}\left( 
\Omega _{\alpha ,0\bar{\beta}}+\Omega_{\mu,0}{}^\mu \delta _{\alpha \bar{\beta}}
+3{g}X\delta _{\alpha \bar{\beta}}\right) ,  \notag \\
H_{\bar{\alpha}\bar{\beta}} &=&2\Omega _{\bar{\alpha},0\bar{\beta}}^{\text{
\ \ \ \ }} \ ,  \label{grav}
\end{eqnarray}
together with the purely geometric constraints

\begin{eqnarray}
\Omega _{\lbrack 0,\alpha ]\beta } &=&0,  \notag \\
\Omega _{0,\mu }^{\text{ \ \ \ \ \ }\mu }-\Omega _{\mu ,0}^{\text{ \ \ \ \ \ 
}\mu }-2{g}X &=&0,  \notag \\
\Omega _{(\alpha ,\mid 0\mid \bar{\beta})} &=&
-{g}X\delta _{\alpha \bar{\beta}} \ ,  \label{geometryone}
\end{eqnarray}
and

\begin{eqnarray}
\Omega _{\alpha ,\mu \nu } &=&0,  \notag \\
\Omega _{\alpha ,\beta }^{\text{ \ \ \ \ }\beta }+\frac{1}{2}\Omega
_{0,0\alpha } &=&0,  \notag \\
\Omega _{\alpha ,\bar{\mu}\bar{\nu}}
-\frac{1}{2}\delta _{\alpha \bar{\mu}}\Omega _{0,0\bar{\nu}}
+\frac{1}{2}\delta _{\alpha \bar{\nu}}\Omega _{0,0\bar{\mu}} &=&0.  \label{geometrytwo}
\end{eqnarray}
The dilatino Killing spinor equation is given by 
\begin{equation}
\left( (F_{MN}^{I}-X^{I}H_{MN})\Gamma ^{MN}-2i\nabla _{M}X^{I}\Gamma
^{M}-4gV_{J}(X^{I}X^{J}-{\frac{3}{2}}Q^{IJ})\right) \epsilon =0
\end{equation}
which, on setting $\epsilon =1$, implies

\begin{eqnarray}
F^{I}{}_{\alpha }{}^{\alpha } &=&X^{I}H_{\alpha }{}^{\alpha },  \notag \\
F^{I}{}_{0\alpha } &=&X^{I}H{}_{0\alpha }-\partial _{\alpha }X^{I},  \notag
\\
F^{I}{}_{\alpha \beta } &=&X^{I}H_{\alpha \beta }  \ , \notag \\
\partial _{0}X^{I} &=&2g(X^{I}V_{J}X^{J}-{\frac{3}{2}}Q^{IJ}V_{J}) \ .
\label{gaugino}
\end{eqnarray}

To proceed, we examine the conditions implied by ({\ref{geometryone}}) and 
({\ref{geometrytwo}}). Define the 1-form $V=\mathbf{e}^{0}$, and introduce a 
$t $ coordinate such that the dual vector field is 
$V=-{\frac{\partial }{\partial t}}$. We also introduce the real coordinates $x^{m}$, 
for $m=1,2,3,4 $. The vielbein is then given by 
\begin{equation}
\mathbf{e}^{0}=dt+\omega _{m}dx^{m},\text{ \ \ }\mathbf{e}^{\alpha }=
\mathbf{e}^{\alpha }{}_{m}dx^{m}\ .
\end{equation}

From (\ref{geometryone}), it can be easily demonstrated that

\begin{eqnarray}
\left( \mathcal{L}_{V}\mathbf{e}^{\alpha }\right) _{\bar{\beta}} &=&0, 
\notag \\
\left( \mathcal{L}_{V}\mathbf{e}^{\alpha }\right) _{\beta } &=&\left( \Omega
_{0,\text{ \ }\beta }^{\text{ \ \ }\alpha }-
\Omega _{\beta ,\text{ \ }0}^{\text{ \ \ }\alpha }+\frac{1}{2}
\left( \Omega _{0,\mu }^{\text{ \ \ \ \ \ }\mu }
+\Omega _{\mu ,}{}^{\mu }{}_{0}\right) \delta ^{\alpha }{}_{\beta}\right) 
-{g}X\delta ^{\alpha }{}_{\beta }.
\end{eqnarray}
The quantity 
\begin{equation}
\Omega _{0,\text{ \ }\beta }^{\text{ \ \ }\alpha }
-\Omega _{\beta ,\text{ \ }0}^{\text{ \ \ }\alpha }+\frac{1}{2}
\left( \Omega _{0,\mu }^{\text{ \ \ \ \ \ }\mu }+\Omega _{\mu ,}{}^{\mu }{}_{0}\right) \delta ^{\alpha }{}_{\beta }
\end{equation}
is anti-Hermitian and traceless (i.e. $\in su(2)$) and as such it can be
gauged away by applying a $SU(2)\subset Spin(4,1)$ gauge transformation to
the Killing spinors, which leaves $1$ invariant. In this gauge, 
\begin{equation}
\mathcal{L}_{V}\mathbf{e}^{\alpha }=-{g}X\mathbf{e}^{\alpha }.
\end{equation}
Define ${\hat{\mathbf{e}}}^{\alpha }$ by 
\begin{equation}
\mathbf{e}^{\alpha }=G{\hat{\mathbf{e}}}^{\alpha }\ ,
\end{equation}
with 
\begin{equation}
\frac{\partial _{t}G}{G}=gX,
\end{equation}
then 
\begin{equation}
\mathcal{L}_{V}{\hat{\mathbf{e}}}^{\alpha }=0.
\end{equation}
In what follows we introduce the base manifold $B$ with the $t$-independent
metric 
\begin{equation}
ds_{B}^{2}=2\delta _{\alpha \bar{\beta}}{\hat{\mathbf{e}}}^{\alpha }
{\hat{\mathbf{e}}}^{\bar{\beta}}\ .
\end{equation}
Let us denote the spin connections on the manifold $B$ by ${\hat{\Omega}}$
and rewrite the conditions in (\ref{geometrytwo}) in terms of ${\hat{\Omega}}
$. The third condition in ({\ref{geometrytwo}}) can be written as 
\begin{equation}
2{\hat{\Omega}}_{\alpha ,\bar{\mu}\bar{\nu}}-2G^{-1}
\big({\tilde{\partial}}_{\hat{\bar{\mu}}}G\delta _{\alpha \bar{\nu}}-
{\tilde{\partial}}_{\hat{\bar{\nu}}}G\delta _{\alpha \bar{\mu}}\big)-G^{-2}
\big(\delta _{\alpha \bar{\mu}}\partial _{t}(G^{2}\omega _{\hat{\bar{\nu}}})
-\delta _{\alpha \bar{\nu}}\partial _{t}(G^{2}\omega _{\hat{\bar{\mu}}})\big)=0.
\end{equation}
Contracting with $\delta ^{\alpha \bar{\nu}}$ we obtain

\begin{equation}
\partial _{t}\left( G^{2}\omega _{{\hat{\bar{\mu}}}}\right) =
-2G^{2}\hat{\Omega}_{\alpha ,\bar{\mu}}^{\text{ \ \ \ \ }\alpha }
+\left( {{\tilde{d}}}G^{2}\right) _{\hat{\bar{\mu}}}  \label{ansar}
\end{equation}
which gives

\begin{equation}
\omega_{{\hat{\bar{\mu}}}}=M\hat{\Omega}_{\alpha ,\bar{\mu}}^{\text{ \ \ \ \
\ }\alpha }+G^{-2} (\mathcal{Q} + {\tilde{d}} \int G^2)_{{\hat{\bar{\mu}}}}
\end{equation}
with

\begin{equation}
\partial _{t} {\mathcal{Q}}_{{\hat{\bar{\mu}}}}=0,\text{ \ \ \ \ \ }M=
-\frac{2}{G^{2}} \int G^{2} \ .
\end{equation}
Therefore we can write

\begin{equation}
\omega =M\mathcal{P}+G^{-2}(\mathcal{Q} + \tilde{d} \int G^2)\ ,
\end{equation}
where 
\begin{equation}
\mathcal{P}\equiv {\hat{\Omega}}_{{\bar{\beta}},\alpha }{}^{\bar{\beta}}{\hat{\mathbf{e}}}^{\alpha }
+{\hat{\Omega}}_{{\beta },{\bar{\alpha}}}{}^{{\beta }}{\hat{\mathbf{e}}}^{\bar{\alpha}}=
\mathcal{P}_{m}dx^{m}\ ,
\label{ppot}
\end{equation}
and $\mathcal{Q}$ are 1-forms on the base manifold $B$ satisfying 
\begin{equation}
\mathcal{L}_{V}\mathcal{Q}=0\ ,\text{ \ \ \ }\mathcal{L}_{V}\mathcal{P}=0.
\end{equation}

The remaining two conditions in (\ref{geometrytwo}) give

\begin{equation}
\hat{\Omega}_{\alpha ,\mu \nu }=0 \ , \label{b1}
\end{equation}
and 
\begin{equation}
\hat{\Omega}_{\alpha ,\mu }^{\text{ \ \ \ \ }\mu }-\hat{\Omega}_{\bar{\mu}
,\alpha }^{\text{ \ \ \ \ \ }\bar{\mu}}=0.  \label{b2}
\end{equation}

It is convenient to define the almost hypercomplex structure

\begin{eqnarray}
\mathbf{J}^{1} &=&{\hat{\mathbf{e}}}^{1}\wedge {\hat{\mathbf{e}}}^{2}
+{\hat{\mathbf{e}}}^{\bar{1}}\wedge {\hat{\mathbf{e}}}^{\bar{2}}\ ,  \notag \\
\mathbf{J}^{2} &=&i{\hat{\mathbf{e}}}^{1}\wedge {\hat{\mathbf{e}}}^{\bar{1}}
+i{\hat{\mathbf{e}}}^{2}\wedge {\hat{\mathbf{e}}}^{\bar{2}}\ ,  \notag \\
\mathbf{J}^{3} &=&-i{\hat{\mathbf{e}}}^{1}\wedge {\hat{\mathbf{e}}}^{2}
+i{\hat{\mathbf{e}}}^{\bar{1}}\wedge {\hat{\mathbf{e}}}^{\bar{2}}\ ,
\end{eqnarray}
where $\mathbf{J}^i$ satisfy the algebra of the imaginary unit quaternions.
The conditions \ ({\ref{b1}}) and ({\ref{b2}}) are then equivalent to

\begin{equation}
d\mathbf{J}^{i}=-2\mathcal{P}\wedge \mathbf{J}^{i}\ ,\qquad i=1,2,3\ ,
\label{hyperhermit}
\end{equation}
where ${\mathcal{P}}$ is given by ({\ref{ppot}}). The condition 
({\ref{hyperhermit}}) implies that the base $B$ is hyper-K\"{a}hler with torsion
(HKT), i.e. 
\begin{equation}
\nabla ^{+}\mathbf{J}^{i}=0\ ,
\end{equation}
where the connection of the covariant derivative $\nabla ^{+}$ is given by 
\begin{equation}
\Gamma ^{(+)}{}^{i}{}_{jk}=\{{}_{jk}^{i}\}+\Theta^{i}{}_{jk}\ ,
\end{equation}
and where $\Theta$ is the torsion 3-form on $B$ given by 
\begin{equation}
\Theta=\star _{4}\mathcal{P}\ ,
\end{equation}
where we take the volume form on the base space $B$ to be $-{\frac{1 }{2}} 
\mathbf{J}^1 \wedge \mathbf{J}^1$, in this convention $\mathbf{J}^i$ are
anti-self-dual. Note that ({\ref{hyperhermit}}) implies that 
\begin{eqnarray}
d {\mathcal{P}}\wedge \mathbf{J}^i=0
\end{eqnarray}
for $i=1,2,3$. Equivalently, the anti-self-dual projection of 
$d {\mathcal{P}}$ vanishes, 
\begin{eqnarray}
(d {\mathcal{P}})^-=0 \ .
\end{eqnarray}

The constraints (\ref{gravitini}) give for the gauge fields

\begin{equation}  \label{vx1a}
A=V_{I}A{}^{I}= -{\frac{2 }{3g}} G^{-1} dG + X \mathbf{e}^0 + {\frac{2 }{3g}}
{\mathcal{P}}
\end{equation}
and 
\begin{eqnarray}  \label{vx1b}
H = d \mathbf{e}^0 + \Psi
\end{eqnarray}
where $\Psi$ is a traceless (1,1) form on $B$, i.e. $\Psi$ is a self-dual
2-form on $B$, with 
\begin{eqnarray}  \label{vx1c}
\Psi = {\frac{4 }{3}} \big( G^{-2} \int G^2 \big) d {\mathcal{P}} -{\frac{2 
}{3}} G^{-2} \big( d {\mathcal{Q}} +2 {\mathcal{Q}} \wedge {\mathcal{P}} 
\big)^+
\end{eqnarray}
where here ${}^+$ denotes the self-dual projection onto the base space $B$.
Next we consider the conditions obtained from ({\ref{gaugino}}). The first
three conditions imply that 
\begin{eqnarray}
F^I= d \big( X^I \mathbf{e}^0 \big) + \Psi^I
\end{eqnarray}
where $\Psi^I$ are closed, $t$-independent self-dual 2-forms on $B$,
satisfying, as a consequence of ({\ref{vx1a}}), 
\begin{eqnarray}  \label{vx2}
V_I \Psi^I = {\frac{2 }{3g}} d {\mathcal{P}}
\end{eqnarray}
and as a consequence of ({\ref{vx1b}}) 
\begin{eqnarray}  \label{vx3}
X_I \Psi^I = \Psi
\end{eqnarray}
where $\Psi$ is given by ({\ref{vx1c}}).

The final condition in ({\ref{gaugino}}) implies that 
\begin{eqnarray}  \label{vx4}
X_I = 2g \big( G^{-2} \int G^2 \big) V_I + G^{-2} Z_I , \qquad \partial_t
Z_I =0
\end{eqnarray}
where $Z_I$ are $t$-independent functions on $B$.

On substituting ({\ref{vx4}}) into ({\ref{vx3}}), and making use of 
({\ref{vx2}}) one obtains 
\begin{equation}
Z_{I}\Psi ^{I}=-{\frac{2}{3}}\big(d{\mathcal{Q}}+2{\mathcal{Q}}\wedge 
{\mathcal{P}}\big)^{+} \ .
\end{equation}

Next, it is convenient to make a co-ordinate transformation to simplify the
solution, and define 
\begin{equation}
u=\int G^{2}.
\end{equation}

The metric, gauge field strengths and scalars are then given by

\begin{eqnarray}
ds^{2} &=&-G^{-4}\big(du-2u{\mathcal{P}}+{\mathcal{Q}}\big)^{2}
+G^{2}ds_{B}^{2},  \notag  \label{ssol1} \\
F^{I} &=&d\bigg(G^{-2}X^{I}\big(du-2u{\mathcal{P}}+{\mathcal{Q}}\big)\bigg)
+\Psi ^{I},  \notag \\
X_{I} &=&G^{-2}\big(2guV_{I}+Z_{I}\big).
\end{eqnarray}
where $ds_{B}^{2}$ is the $u$-independent metric on a (strong) HKT manifold 
$B$. ${\mathcal{P}}$ is a $u$-independent 1-form on $B$ satisfying 
({\ref{hyperhermit}}) and $d{\mathcal{P}}$ is a self-dual 2-form on $B$. 
${\mathcal{Q}}$ is another $u$-independent 1-form on $B$, $Z_{I}$ are 
$u$-independent functions on $B$, and $\Psi ^{I}$ are closed self-dual, 
$u$-independent 2-forms on $B$ satisfying 
\begin{equation}
V_{I}\Psi ^{I}={\frac{2}{3g}}d{\mathcal{P}}  \label{ssol1v1}
\end{equation}
and 
\begin{equation}
Z_{I}\Psi ^{I}=-{\frac{2}{3}}\big(d{\mathcal{Q}}+2{\mathcal{Q}}\wedge 
{\mathcal{P}}\big)^{+}.  \label{ssol1v2}
\end{equation}

It remains to consider the gauge field equations; one finds, after some
computation, that 
\begin{equation}
{\hat{\nabla}}^{i}\big(-{\frac{3}{2}}dZ_{I}+3Z_{I}{\mathcal{P}}
+3gV_{I}{\mathcal{Q}}\big)_{i}+{\frac{1}{8}}C_{IJK}\Psi _{ij}^{J}\Psi ^{Kij}=0
\label{geq1}
\end{equation}
where ${\hat{\nabla}}$ denotes the Levi-Civita connection of $B$, and here
all indices are frame indices on $B$. We remark that, as a consequence of
the integrability conditions examined in Appendix B of \cite{gshalf}, the
Killing spinor equations, together with the gauge field equations and
Bianchi identity are sufficient to imply that the Einstein and the scalar
field equations hold automatically, without any further constraint.

Note that the solution ({\ref{ssol1}}) together with the conditions in 
({\ref{ssol1v1}}), ({\ref{ssol1v2}}) and ({\ref{geq1}}) are invariant under the
conformal re-scaling 
\begin{equation}
ds_{B}^{2}=e^{-2h}ds_{B}^{\prime 2}  \label{conf1}
\end{equation}
where $h$ is a $u$-independent function, together with the re-definitions 
\begin{equation}
u=e^{2h}u^{\prime },\qquad {\mathcal{P}}={\mathcal{P}}^{\prime }+dh,
\qquad {\mathcal{Q}}=e^{2h}{\mathcal{Q}}^{\prime },\qquad Z_{I}=e^{2h}Z_{I}^{\prime }, 
\qquad G= e^h G^\prime \ .
  \label{conf2}
\end{equation}
A HKT manifold is called \textit{strong} HKT if the associated torsion 
$\Theta $ is closed, or equivalently 
\begin{equation}
d\star _{4}{\mathcal{P}}=0
\end{equation}
where $\star _{4}$ denotes the Hodge dual on $B$. For the solutions under
consideration here, by making an appropriate conformal transformation as
described above, one can without loss of generality take $B$ to be a strong
HKT manifold.

In order to recover the solutions for the minimal theory determined in 
\cite{hkt5}, one sets 
\begin{equation}
C_{111}={\frac{2}{\sqrt{3}}},\qquad X^{1}=\sqrt{3},\qquad X_{1}=
{\frac{1}{\sqrt{3}}}
\end{equation}
and hence we set 
\begin{equation}
V_{1}={\frac{1}{\sqrt{3}}},\qquad Z_{1}=0
\end{equation}
In addition, one has 
\begin{equation}
G=e^{gt},\qquad g=-{\frac{\chi }{2\sqrt{3}}},\qquad \Psi ^{1}=
-{\frac{4}{\chi }}d{\mathcal{P}},\qquad F^{1}=2F
\end{equation}
where $F$ is the Maxwell field strength of the minimal theory.

It is also useful to consider a co-ordinate transformation of the form 
\begin{equation}
u^{\prime }=u-\Theta
\end{equation}
where the function $\Theta $ does not depend on $u$, and set 
\begin{equation}
{\mathcal{Q}}^{\prime }={\mathcal{Q}}-2\Theta {\mathcal{P}}+d\Theta ,\qquad
Z_{I}^{\prime }=Z_{I}+2g\Theta V_{I} \ .
\end{equation}
Under these transformations, the solution given in ({\ref{ssol1}}), ({\ref%
{ssol1v1}}), ({\ref{ssol1v2}}) , together with the gauge equations ({\ref%
{geq1}}) are invariant. It is clear that one can always choose the function $%
\Theta $ such that 
\begin{equation}
d\star _{4}{\mathcal{Q}}^{\prime }=0
\end{equation}
and one can therefore work in a gauge for which \textit{both} ${\mathcal{P}}$
and ${\mathcal{Q}}$ are co-closed. It should however be noted that the gauge
in which ${\mathcal{Q}}$ is co-closed is not the same gauge in which the
solutions to the minimal theory are constructed as described in \cite{hkt5};
this is because the minimal theory gauge has $Z_{1}=0$ and $G$ is a function
only of $t$. One cannot in general make a gauge transformation of the form
described above and keep $Z_{1}^{\prime }=0$ as well. In what follows it
will be most convenient to work with the gauge choice for which 
\begin{equation}
d\star _{4}{\mathcal{P}}=d\star _{4}{\mathcal{Q}}=0\ .
\end{equation}

\section{Solutions with a tri-holomorphic isometry}

It is straightforward to analyse the case when the base manifold $B$ is
strong HKT and admits a tri-holomorphic isometry, which we denote by 
${\frac{\partial }{\partial x^{5}}}$, and we take this isometry to be a symmetry of
the full solution. Such base spaces have been classified in \cite{gaudtod, papadmon,
chavetodvalent}, and the metric on $B$ is given by 
\begin{equation}
ds_{B}^{2}=W^{-1}\big(dx^{5}+\varphi \big)^{2}+Wds_{E}^{2}  \label{tri1}
\end{equation}
where $E$ is a constrained 3-dimensional Einstein-Weyl geometry, consisting
of a $x^{5}$-independent 3-metric $\gamma _{ij}$, a $x^{5}$-independent
1-form $\alpha $ on $E$, and an $x^{5}$-independent scalar $\alpha _{0}$ on 
$E$, satisfying 
\begin{equation}
\star _{E}d\alpha =-d\alpha _{0}-\alpha _{0}\alpha ,\qquad d\star _{E}\alpha
=0
\end{equation}
where $\star _{E}$ denotes the Hodge dual on $E$, and the Ricci tensor of $E$
satisfies 
\begin{equation}
{}^{(E)}R_{ij}+\nabla _{(i}\alpha _{j)}+\alpha _{i}\alpha _{j}=\gamma _{ij}
({\frac{1}{2}}\alpha _{0}^{2}+\alpha ^{\ell }\alpha _{\ell })
\end{equation}
where here $\nabla $ denotes the Levi-Civita connection of $E$, and $\varphi 
$ is a $x^{5}$ independent 1-form on $E$ satisfying 
\begin{equation}
\star _{E}d\varphi =dW+W\alpha
\end{equation}
and the function $W$ does not depend on $x^{5}$. The volume form of $B$, 
$\epsilon _{B}$, and the volume form of $E$, $d\mathrm{vol}_{E}$ are related
by 
\begin{equation}
\epsilon _{B}=W(dx^{5}+\varphi )\wedge d\mathrm{vol}_{E}.
\end{equation}
The torsion of $B$ is determined by ${\mathcal{P}}$, with 
\begin{equation}
{\mathcal{P}}=-{\frac{\alpha _{0}}{2W}}(dx^{5}+\varphi )-{\frac{1}{2}}\alpha
\end{equation}
which is co-closed as a consequence of the previous conditions. We further
remark that the functions $W$, $\alpha _{0}$ satisfy 
\begin{equation}
(\Delta _{E}+\alpha ^{i}\nabla _{i})W=(\Delta _{E}+\alpha ^{i}\nabla
_{i})\alpha _{0}=0
\end{equation}
where $\Delta _{E}$ is the Laplacian on $E$.

To proceed further with the analysis, note that self-duality of $\Psi ^{I}$,
together with the requirement that $d\Psi ^{I}=0$, are equivalent to 
\begin{equation}
\Psi ^{I}=-{\frac{1}{2}}(dx^{5}+\varphi )\wedge d\big(W^{-1}K^{I}\big)-
{\frac{1}{2}}W\star _{E}d\big(W^{-1}K^{I}\big)
\end{equation}
where $K^{I}$ are $x^{5}$-independent functions on $E$ satisfying 
\begin{equation}
(\Delta _{E}+\alpha ^{i}\nabla _{i})K^{I}=0.
\end{equation}
The condition ({\ref{ssol1v1}}) constrains the $K^{I}$ via 
\begin{equation}
V_{I}K^{I}={\frac{1}{3g}}\alpha _{0}+kW  \label{ssol3a2}
\end{equation}
for constant $k$. Next, it is straightforward to solve the gauge equation 
({\ref{geq1}}) to find 
\begin{equation}
Z_{I}={\frac{1}{24}}C_{IJK}W^{-1}K^{J}K^{K}+L_{I}
\end{equation}
where $L_{I}$ are $x^{5}$-independent functions on $E$ satisfying 
\begin{equation}
(\Delta _{E}+\alpha ^{i}\nabla _{i})L_{I}=0.
\end{equation}
Finally we solve for the 1-form ${\mathcal{Q}}$. We decompose this 1-form as 
\begin{equation}
{\mathcal{Q}}=Q_{5}(dx^{5}+\chi )+{\tilde{Q}}
\end{equation}
where the function $Q_{5}$ does not depend on $x^{5}$, and ${\tilde{Q}}$ is
a $x^{5}$-independent 1-form on $E$. The condition $d\star _{B}{\mathcal{Q}}
=0$ implies that 
\begin{equation}
d\star _{E}\tilde{Q}=0
\end{equation}
and the condition ({\ref{ssol1v2}}), after some manipulation, implies that 
\begin{equation}
Q_{5}=-{\frac{1}{48}}W^{-2}C_{IJK}K^{I}K^{J}K^{K}-{\frac{3}{4}}
W^{-1}L_{I}K^{I}+M
\end{equation}
where $M$ is a $x^{5}$-independent function on $E$ satisfying 
\begin{equation}
(\Delta _{E}+\alpha ^{i}\nabla _{i})M=0
\end{equation}
and ${\tilde{Q}}$ also must satisfy 
\begin{equation}
d{\tilde{Q}}+\alpha \wedge {\tilde{Q}}+\alpha _{0}\star _{E}{\tilde{Q}}
=W\star _{E}dM-M\star _{E}dW-{\frac{3}{4}}\big(K^{I}\star
_{E}dL_{I}-L_{I}\star _{E}dK^{I}\big).
\end{equation}

\section{Solutions with a Conformally Hyper-K\"ahler base}

Suppose that the base space $B$ is conformally hyper-K\"{a}hler. Then 
${\mathcal{P}}$ is closed, and using the conformal transformation described in
({\ref{conf1}}) and ({\ref{conf2}}), one can without loss of generality set 
${\mathcal{P}}=0$, i.e. one can take $B$ to be a hyper-K\"{a}hler manifold,
which we denote by $HK$. We shall also work in a gauge for which 
${\mathcal{Q}}$ is co-closed on $HK$, as described previously. Hence the solution can be
written as 
\begin{eqnarray}
ds^{2} &=&-G^{-4}\big(du+{\mathcal{Q}}\big)^{2}+G^{2}ds_{HK}^{2},  \notag
\label{ssol1b} \\
F^{I} &=&d\bigg(G^{-2}X^{I}\big(du+{\mathcal{Q}}\big)\bigg)+\Psi ^{I}, 
\notag \\
X_{I} &=&G^{-2}\big(2guV_{I}+Z_{I}\big),
\end{eqnarray}
where $ds_{HK}^{2}$ is the $u$-independent metric on a hyper-K\"{a}hler
manifold $HK$, ${\mathcal{Q}}$ is a $u$-independent 1-form on $HK$, $Z_{I}$
are $u$-independent functions on $HK$, and $\Psi ^{I}$ are self-dual, 
$u$-independent 2-forms on $HK$ satisfying 
\begin{equation}
d\Psi ^{I}=0  \label{ssol2a}
\end{equation}
and 
\begin{equation}
V_{I}\Psi ^{I}=0  \label{ssol2b}
\end{equation}
and 
\begin{equation}
Z_{I}\Psi ^{I}=-{\frac{2}{3}}\big(d{\mathcal{Q}}\big)^{+}  \label{ssol2c}
\end{equation}
and 
\begin{equation}
d\star _{4}{\mathcal{Q}}=0  \label{ssol2d}
\end{equation}
and 
\begin{equation}
d\star _{4}dZ_{I}+{\frac{1}{6}}C_{IJK}\Psi ^{J}\wedge \Psi ^{K}=0 \ .
\label{ssol2e}
\end{equation}

In order to recover the special case for which the base space is 
hyper-K\"{a}hler with a triholomorphic isometry, i.e. a Gibbons-Hawking manifold, for
which the triholomorphic isometry is a symmetry of the full solution, one
takes the analysis of the previous section and sets $E=\mathbb{R}^{3}$, 
$\alpha _{0}=0$, $\alpha =0$, with $W=H$ where $H$ is a harmonic function on 
$\mathbb{R}^{3}$ and $\varphi $ is a $x^{5}$-independent 1-form on 
$\mathbb{R}^{3}$ satisfying 
\begin{equation}
d\varphi =\star _{\mathbb{R}^{3}}dH
\end{equation}
and the remaining functions $K^{I}$, $L_{I}$, $M$ which are used in the
construction of the solution are also harmonic functions on $\mathbb{R}^{3}$. 
We remark that one can also allow $Z_{I}$ to depend linearly on $x^{5}$ by
taking
\begin{equation}
Z_{I}={\frac{1}{24}}C_{IJK}H^{-1}K^{J}K^{K}+L_{I}+cV_{I}x^{5}
\end{equation}
for constant $c$, with $\Psi ^{I}$, ${\mathcal{Q}}$ unchanged (and $H$, 
$K^{I}$, $L_{I}$, $M$ still $x^{5}$-independent). It is straightforward to
show that adding such a term linear in $x^{5}$ to $Z_{I}$ does not give any
contribution to the LHS of the conditions ({\ref{ssol2c}}) and ({\ref{ssol2e}}).

The black hole solutions found in \cite{liusabra, Klemm:2000gh,
Behrndt:2003cx} are a special case of the solutions found here, for which
all the harmonic functions depend only on $r$, and hence have poles only at 
$r=0$.

\section{Final Remarks}

In this paper we have studied timelike solutions admitting Killing spinors
of five dimensional de Sitter supergravity with Abelian vector multiplets.
The four dimensional base space of these solutions was found to be given by
a four dimensional HKT geometry. In our present work we have also described two special classes of solution.
First, we considered the case when the HKT manifold admits a tri-holomorphic Killing vector field.
Then we considered the case for which the HKT manifold is conformally hyper-K\"ahler.
The conformally hyper-K\"{a}hler class
of solutions includes all previously constructed solutions in the literature
as special cases. 
It would be of great interest to construct new solutions in the
non-conformally hyper-K\"{a}hler case, as these  might be of
relevance to black hole physics and cosmology. It would also be particularly
interesting to determine whether there exist regular (pseudo) supersymmetric
black ring solutions in de Sitter supergravity. Finally, a continuation of
our present work is to study the solutions of the null case for the theories
considered here and possibly generalising these results to de Sitter
supergravity in other dimensions. Work along these directions is in progress.

\section*{Acknowledgments}

The work of WS was supported in part by the National Science Foundation
under grant number PHY-0903134. JG is supported by the EPSRC grant
EP/F069774/1.

\renewcommand{\theequation}{A-\arabic{equation}} 
\setcounter{equation}{0} 


\begin{thebibliography}{99}
\bibitem{Zumino:1979et} B.~Zumino, 
\textit{Supersymmetry And Kahler Manifolds},\ Phys.\ Lett. \textbf{B87} (1979) 203.

\bibitem{AlvarezGaume:1981hm} L.~Alvarez-Gaume and D.~Z.~Freedman, 
\textit{Geometrical Structure And Ultraviolet Finiteness In The Supersymmetric Sigma
Model},\ Commun.\ Math.\ Phys.\ \textbf{80} (1981) 443.

\bibitem{Witten:1983ar} E.~Witten, \textit{Nonabelian bosonization in two
dimensions}, Commun.\ Math.\ Phys.\ \textbf{92} (1984) 455.

\bibitem{Curtright:1984dz} T.~L.~Curtright and C.~K.~Zachos, Geometry, 
\textit{Topology And Supersymmetry In Nonlinear Models}, Phys.\ Rev.\ Lett.\ 
\textbf{53} (1984) 1799.

\bibitem{Braaten:1985is} E.~Braaten, T.~L.~Curtright and C.~K.~Zachos, 
\textit{Torsion And Geometrostasis In Nonlinear Sigma Models}, Nucl.\ Phys.\
\textbf{B260} (1985) 630.

\bibitem{howepaphkt} P. S. Howe and G. Papadopoulos, 
\textit{Ultraviolet Behavior Of Two-Dimensional Supersymmetric Nonlinear
Sigma Models}, Nucl. Phys. \textbf{\ B289} (1987) 264; \textit{Further
Remarks On The Geometry Of Two-Dimensional Nonlinear Sigma Models}, Class.
Quant. Grav. \textbf{5} (1988) 1647.

\bibitem{hullhkt} C. M. Hull, \textit{Lectures on Nonlinear Sigma Models and
Strings}, Lectures given in the Super Field Theories workshop, Vancouver
Canada (1986).

\bibitem{hull:1984}
S. J. Gates, C. M. Hull and M. Rocek, \textit{Twisted Multiplets and New 
Supersymmetric Nonlinear Sigma Models,}
Nucl. Phys. {\bf{B248}} (1984) 157.

\bibitem{hull:1986}
C. M. Hull, A. Karlhede, U. Lindstrom and M. Rocek,
\textit{Nonlinear Sigma Models and their Gauging In and Out of Superspace,}
Nucl. Phys. {\bf{B266}} (1986) 1.

\bibitem{Candelas:1985en} P.~Candelas, G.~T.~Horowitz, A.~Strominger and
E.~Witten, \textit{Vacuum Configurations For Superstrings}, \ Nucl.\ Phys.\ 
\textbf{B258} (1985) 46.

\bibitem{gibbonshkt} G. W. Gibbons, G. Papadopoulos and K. Stelle, 
\textit{HKT and OKT geometries on soliton black hole moduli spaces}, Nucl. Phys. 
\textbf{\ B508} (1997) 623; [arXiv:hep-th/9706207].

\bibitem{multihkt} J. Gutowski and G. Papadopoulos, \textit{The Dynamics of
very special black holes}, Phys. Lett. \textbf{\ B472} (2000) 45;
[arXiv:hep-th/9910022].

\bibitem{Gauntlett:2002nw} J.~P.~Gauntlett, J.~B.~Gutowski, C.~M.~Hull,
S.~Pakis and H.~S.~Reall, \textit{All supersymmetric solutions of minimal
supergravity in five dimensions},\ Class.\ Quant.\ Grav.\ \textbf{20} (2003)
4587; [arXiv:hep-th/0209114].

\bibitem{Gauntlett:2003fk} J.~P.~Gauntlett and J.~B.~Gutowski,
\textit{All supersymmetric solutions of minimal gauged
supergravity in five dimensions}, Phys.\ Rev.\  \textbf{D68} (2003) 105009 [Erratum-ibid.\  \textbf{D70} (2004) 089901]; [arXiv:hep-th/0304064]. J. B.
Gutowski and W. A. Sabra, \textit{General supersymmetric solutions of
five-dimensional supergravity}, JHEP \textbf{10} $(2005)$ 039;
[arXiv:hep-th/0505185]. 

\bibitem{hkt5} J. Grover, J. B. Gutowski, Carlos A. R. Herdeiro and W. A.
Sabra, \textit{Five Dimensional Minimal Supergravities and Four Dimensional
Complex Geometries}, Contribution to the Proceedings of the Spanish
Relativity Meeting 2008 in Salamanca, Spain, [arXiv:0901.4066 (hep-th)]. 
J. Grover, J. B. Gutowski, Carlos A. R. Herdeiro and W. A. Sabra, HKT \textit{Geometry and
de Sitter Supergravity}, Nucl. Phys. \textbf{\ B809} (2009) 406; [arXiv:0806.2626 (hep-th)].

\bibitem{gtnull5d} J. Grover, Jan B. Gutowski, Carlos A.R. Herdeiro, P.
Meessen, A. Palomo-Lozano and W. A. Sabra, \textit{Gauduchon-Tod structures,
Sim holonomy and De Sitter supergravity}, JHEP \textbf{07} (2009) 069;
[arXiv:0905.3047 (hep-th)].

\bibitem{gaudtod} P. Gauduchon and K. P. Tod, \textit{Hyper-Hermitian metric
with symmetry}, Jour. Geom. Phys. \textbf{25} (1998) 291.

\bibitem{maeda} Masato Nozawa and Kei-ichi Maeda, {\textit{Cosmological rotating
black holes in five-dimensional fake supergravity}}; [arXiv:1009.3688 (hep-th)].

\bibitem{gunaydin} M. Gunaydin, G. Sierra and P.K. Townsend, 
\textit{Gauging the d = 5 Maxwell-Einstein Supergravity Theories: More on Jordan
Algebras}, Nucl. Phys.  \textbf{B253} (1985) 573.


\bibitem{papadopd11} J. Gillard, U. Gran and G. Papadopoulos, \textit{The
Spinorial geometry of supersymmetric backgrounds}, Class. Quant. Grav. 
\textbf{22} (2005) 1033; [arXiv:hep-th/0410155].

\bibitem{papadopiib} U. Gran, J. Gutowski and G. Papadopoulos, \textit{The
Spinorial geometry of supersymmetric IIb backgrounds}, Class. Quant. Grav. 
\textbf{22} (2005) 2453; [arXiv:hep-th/0501177].



\bibitem{gshalf} J. B. Gutowski and W. A. Sabra, \textit{\
Half-Supersymmetric Solutions in Five-Dimensional Supergravity,} JHEP 
\textbf{12} (2007) 025, Erratum-ibid.1004 (2010) 042;
[arXiv:0706.3147 (hep-th)].


\bibitem{papadmon} G. Papadopoulos,
\textit{Elliptic monopoles and (4,0) supersymmetric sigma models with torsion,}
Phys. Lett. {\bf{B356}} (1995) 249; 
[arXiv:hep-th/9505119].



\bibitem{chavetodvalent} T. Chave, K. P. Tod and G. Valent, (\textit{4,0)
and (4,4) sigma models with a tri-holomorphic Killing vector}, 
Phys. Lett. \textbf{\ B383} (1996) 262.

\bibitem{liusabra} J. T. Liu and W. A. Sabra, \textit{Multicentered black
holes in gauged D = 5 supergravity}, Phys. Lett. \textbf{B498} (2001) 123;
[arXiv:hep-th/0010025].

\bibitem{Klemm:2000gh} D.~Klemm and W.~A.~Sabra, \textit{General (anti-)de
Sitter black holes in five dimensions}, JHEP \textbf{02} (2001) 031;
[arXiv:hep-th/0011016].

\bibitem{Behrndt:2003cx} K.~Behrndt and M.~Cvetic, \textit{Time-dependent
backgrounds from supergravity with gauged non-compact R-symmetry}, 
Class.\ Quant.\ Grav.\ \textbf{20} (2003) 4177; [arXiv:hep-th/0303266].




\end{thebibliography}
\end{document}